\documentclass[fleqn,usenatbib]{mnras}

\usepackage{graphicx}
\usepackage{amsmath}
\usepackage[dvipsnames]{xcolor}
\usepackage{amssymb}
\usepackage{newtxtext}
\usepackage[varvw]{newtxmath}
\usepackage{ulem}
\usepackage{soul}
\usepackage{siunitx}
\usepackage{macros_vdb}
\usepackage{physics}
\usepackage{tikz}
\usetikzlibrary{shapes,arrows}

\defcitealias{Navarro.etal.97}{NFW}
\defcitealias{MN1975}{MN}
\defcitealias{Green2021}{G21}
\defcitealias{Green.vdBosch.19}{GB19}

\graphicspath{{./figures/}}

\title[The impact of a disc potential on subhaloes]{\texttt{SatGen} -- II. Assessing the impact of a disc potential on subhalo populations}

\author[S. B. Green, F. C. van den Bosch, and  F. Jiang]{%
Sheridan~B.~Green,$^{1}$\thanks{E-mail: \href{mailto:sheridan.green@yale.edu}{sheridan.green@yale.edu} (SBG)}\thanks{NSF Graduate Research Fellow} Frank~C.~van den Bosch,$^{1,2}$ and Fangzhou Jiang$^{3,4}$\thanks{Troesh Scholar}
\vspace*{8pt}
\\
$^{1}$Department of Physics, Yale University, P.O. Box 208120, New Haven, CT 06520-8120\\
$^{2}$Department of Astronomy, Yale University, P.O. Box 208101, New Haven, CT 06520-8101\\
$^{3}$TAPIR, California Institute of Technology, Pasadena, CA 91125\\
$^{4}$Carnegie Observatories, 813 Santa Barbara Street, Pasadena, CA 91101\\
}

\date{}

\pubyear{2021}

\begin{document}
\label{firstpage}
\pagerange{\pageref{firstpage}--\pageref{lastpage}}
\maketitle

\begin{abstract}
  The demographics of dark matter substructure depend sensitively on the nature of dark matter. Optimally leveraging this probe requires accurate theoretical predictions regarding the abundance of subhaloes. These predictions are hampered by artificial disruption in numerical simulations, by large halo-to-halo variance, and by the fact that the results depend on the baryonic physics of galaxy formation. In particular, numerical simulations have shown that the formation of a central disc can drastically reduce the abundance of substructure compared to a dark matter-only simulation, which has been attributed to enhanced destruction of substructure due to disc shocking. We examine the impact of discs on substructure using the semi-analytical subhalo model \texttt{SatGen}, which accurately models the tidal evolution of substructure free of the numerical disruption that still hampers $N$-body simulations. Using a sample of 10,000 merger trees of Milky Way-like haloes, we study the demographics of subhaloes that are evolved under a range of composite halo--disc potentials with unprecedented statistical power. We find that the overall subhalo abundance is relatively insensitive to properties of the disc aside from its total mass. For a disc that contains 5\% of $M_\mathrm{vir}$, the mean subhalo abundance within $r_\mathrm{vir}$ is suppressed by ${\lesssim}10\%$ relative to the no-disc case, a difference that is dwarfed by halo-to-halo variance. For the same disc mass, the abundance of subhaloes within 50 kpc is reduced by ${\sim}30\%$. We argue that the disc mainly drives excess mass loss for subhaloes with small pericentric radii and that the impact of disc shocking is negligible.
\end{abstract}


\section{Introduction}\label{sec:intro}

The substructure present in dark matter (DM) haloes is the outcome of a hierarchical assembly process combined with tidal and impulsive forces that work to dissolve it. Since the particle nature of DM impacts the mass function and density profiles of DM haloes, it also affects the demographics of its substructure. For example, if DM is ``warm'', the abundance of low-mass subhaloes is suppressed relative to that which is predicted for cold dark matter \citep[e.g.,][]{Lovell.etal.14, Bose.etal.17}. If DM undergoes significant self-interaction, or is an ultra-light boson, the inner halo density profile becomes cored \citep[e.g.,][]{Kaplinghat.etal.16, Robles2017, Burkert2020} and is less resilient to tidal forces \citep[][]{Penarrubia.etal.10}, resulting in an overall suppression of substructure. This powerful potential to place constraints on DM has prompted various observational attempts to quantify the abundance of DM substructure, including searches for gaps in stellar streams \citep[e.g.,][]{Erkal.etal.16, Banik.etal.19, Bonaca2020}, measurements of gravitational lensing distortions \citep[e.g.,][]{Vegetti.etal.14, Hezaveh2016, Nierenberg.etal.17}, indirect detection studies that search for DM annihilation signals \citep[e.g.,][]{Stref.Lavalle.17, Somalwar2021}, and measurements of the abundance of satellite galaxies \citep[via the galaxy--halo connection; e.g.,][]{Nadler2020b}.

In order to fully leverage these observations to constrain DM microphysics, it is prudent that we are able to accurately predict subhalo abundances for the different DM models. Arguably, the best way to account for all the relevant, strongly non-linear physical processes is to use full cosmo-hydrodynamical simulations of galaxy formation \citep[e.g.,][]{Wetzel.etal.16, Pillepich2018} at a resolution sufficient to resolve substructure in the relevant mass range. Unfortunately, the computationally demanding nature of such simulations as well as the uncertainties related to sub-grid physics modeling represent significant roadblocks for their use in such a task. As a consequence, DM-only cosmological simulations \citep[e.g.,][]{Springel.etal.08, Klypin.etal.11} are typically used as a less expensive alternative. However, these simulation-based approaches are still adversely impacted by artificial subhalo disruption and limited mass resolution \citep{vdBosch.17, vdBosch.etal.2018b, Green2021}. Semi-analytical models \citep[SAMs; e.g.,][]{Taylor.Babul.04, Zentner.etal.05, Jiang.vdBosch.16, Jiang2021} provide attractive alternatives for predicting the substructure abundance in a manner that is both computationally efficient and insensitive to the particular numerical limitations of $N$-body simulations.

Unfortunately, DM-only simulations and SAMs typically do not account for the impact of baryons on the subhalo population, which can be quite important. For example, the aforementioned observational probes are most sensitive to the inner halo, where the central galaxy significantly influences the host potential. Several studies have demonstrated that a central galactic disc suppresses the overall subhalo abundance. For example, \citet{DOnghia2010} grew an analytical disc potential in a high-resolution cosmological zoom-in simulation of a Milky Way-like (MW) halo and showed that substructure in the inner regions of the halo is efficiently destroyed, which they ascribed to disc shocking. More recently, \citet{Garrison-Kimmel.etal.17} found that the suppression in subhalo abundance seen in a full physics simulation relative to a DM-only realization of the same halo can be reproduced by simply embedding a disc potential within the DM-only halo. Both \citet{Penarrubia.etal.10} and \citet{Errani2017} used idealized simulations to examine the impact of a central disc on the abundance of subhaloes, confirming once more that the presence of a disc can significantly deplete the subhalo population, especially towards the centre of the halo.

Since a disc potential drives additional subhalo mass loss, its presence must be properly accounted for in any successful substructure modeling endeavor. However, to date, no study has been able to assess the impact of the disc on subhalo populations in a statistically meaningful way. Recently, we introduced \texttt{SatGen} \citep{Jiang2021}, a SAM framework that can rapidly generate random substructure realizations, thereby enabling a robust treatment of the halo-to-halo variance. As shown in \citet{Jiang.vdBosch.17}, this variance can be very large and is strongly correlated with the formation time and concentration of the host halo \citep[see also][]{Zentner.etal.05, Giocoli.etal.10}. Furthermore, \texttt{SatGen} can be used to isolate the influence of the disc from assembly history variation by studying how subhaloes from the same merger tree evolve under different host potentials. Due to its speed, \texttt{SatGen} is also ideal for assessing how sensitive the subhalo statistics are to parameters of the disc model via sweeps of the parameter space.

In this paper, we use \texttt{SatGen} to investigate the differential impact of a galactic disc potential on the subhalo populations of Milky Way-like haloes. We initially explored the influence of a disc in \citet{Jiang2021} --- here, we build upon this pilot study by greatly boosting the size of our halo sample, exploring a wide range of disc models, and incorporating a more sophisticated subhalo tidal evolution model. While our findings are in good agreement with the simulation results of \citet{Errani2017} and \citet{Garrison-Kimmel.etal.17}, our ability to study a large halo sample and, thus, probe the halo-to-halo variance sheds new light on the statistical relevance of these results. We track individual subhaloes and illustrate how their masses are altered due to an embedded central disc. We also search for the presence of a disc-driven angular bias in the spatial distribution of subhaloes, as well as show that the overall subhalo abundance is relatively insensitive to the size and growth history of the disc and is only affected by the disc mass. This manuscript is organized as follows. In Section~\ref{sec:methods}, we first provide an overview of our semi-analytical modeling framework. The results are presented in Section~\ref{sec:results}, which is followed by a detailed discussion (Section~\ref{sec:discuss}) as to whether ``disc shocking'' or enhanced tidal stripping serves as the dominant disc-driven subhalo depletion mechanism. Finally, in Section~\ref{sec:summary}, we summarize our findings and motivate future work.

Throughout this work, the halo mass is defined as the mass enclosed within the virial radius, $r_\mathrm{vir}$, inside of which the mean density is equal to  $\Delta_\mathrm{vir}(z)$ times the critical density. For the $\Lambda$ cold dark matter ($\Lambda$CDM) cosmology that we adopt (${h=0.7}$, $\Omega_\rmm = 0.3$, $\Omega_\Lambda = 0.7$, $\Omega_\rmb = 0.0465$, $\sigma_8 = 0.8$, $n_\rms = 1.0$), $\Delta_\mathrm{vir}(z = 0) \approx 100$ and is otherwise well-described by the fitting formula presented by \citet{Bryan.Norman.98}. Throughout, we use $m$ and $M$ to denote subhalo and host halo masses, respectively. We use $l$ and $r$ to reference subhalo- and host halo-centric radii, respectively. The base-10 logarithm is denoted by $\log$ and the natural logarithm is denoted by $\ln$.

\section{Semi-analytical methods}\label{sec:methods}

This study employs the \texttt{SatGen} semi-analytical modeling framework that is presented by \citet{Jiang2021}. In particular, we use the model of subhalo tidal evolution recently developed by \citet{Green.vdBosch.19} and \citet{Green2021}. This model has been calibrated using the \textit{Dynamical Aspects of Subhaloes} idealized simulation library \citep[hereafter \textit{DASH};][]{Ogiya2019} to accurately reproduce the bound mass and density profiles of simulated $N$-body subhaloes as they orbit within an analytical host potential. We refer the reader to these papers for a comprehensive description of the model. In short, \texttt{SatGen} combines prescriptions for (i) analytical halo merger trees \citep{Parkinson.etal.08}, (ii) subhalo orbit initialization \citep{Li2020}, (iii) orbit integration, including dynamical friction \citep{Chandrasekhar.43}, (iv) density profile evolution \citep[][]{Green.vdBosch.19}, and (v) tidal mass-loss \citep{Green2021} in order to generate subhalo catalogs (which include both mass and position information) for ensembles of host halo realizations.

Both the host halo and the initial subhaloes (i.e., at infall) are assumed to have \citet[][hereafter \citetalias{Navarro.etal.97}]{Navarro.etal.97} density profiles with concentrations computed via the model of \citet{Zhao2009}. Each subhalo is integrated along its orbit and experiences tidal mass loss, which is given by
\begin{equation}\label{eqn:massloss}
  \frac{\Delta m}{\Delta t} = - \alpha \frac{m(>l_\rmt)}{t_\mathrm{char}} \,.
\end{equation}
Here, $t_\mathrm{char}$ is the characteristic orbital time of the subhalo, $m(>l)$ is the subhalo mass that lies outside of radius $l$, with $l_\rmt$ denoting the instantaneous tidal radius (defined below), and $\alpha$ is a calibrated parameter that controls the stripping efficiency. Motivated by the work of \citet{Hayashi.etal.03} and \citet{Penarrubia2008}, the density profiles of stripped subhaloes are modelled according to
\begin{equation}\label{eqn:eshdp}
 \rho(l,t) = H(l|\, f_\rmb(t), c_\mathrm{vir,s}) \, \rho(l,t_{\rm acc}),
\end{equation}
where $f_\rmb(t)$ is the bound mass fraction of the subhalo, $c_\mathrm{vir,s}$ is the concentration of the subhalo \textit{at accretion}, and $t_\mathrm{acc}$ denotes the time of accretion. For the `transfer function', $H(l)$, we use the model of \citet{Green.vdBosch.19}, which has been carefully calibrated against the \textit{DASH} simulations.

The galactic disc, when included, is positioned at the centre of the host halo and modeled with the axisymmetric \citet[][hereafter \citetalias{MN1975}]{MN1975} density profile, which has three parameters: (i) the radial scale length, $a_\rmd$, (ii) vertical scale height, $b_\rmd$, and (iii) mass, $M_\rmd$. We write $a_\rmd = f_a [M_\mathrm{vir}(z) / M_0]^{\beta_a} r_\mathrm{vir,0}$, $M_\rmd = f_M [M_\mathrm{vir}(z) / M_0]^{\beta_M} M_0$, and set $b_\rmd$ to be a fixed fraction of $a_\rmd$. Here, $M_\mathrm{vir}(z)$ is the mass accretion history of the host halo, $M_0 = M_\mathrm{vir}(z=0)$, and $r_\mathrm{vir,0}$ is the virial radius of the host at $z=0$. In this work, we restrict ourselves to host haloes that reach a virial mass of $M_0 = 10^{12}\, h^{-1} M_\odot$ at $z=0$, which corresponds to $r_\mathrm{vir,0} \approx 290$ kpc. Our fiducial, Milky Way-like disc is described by $f_a = 0.0125$, $b_\rmd / a_\rmd = 0.08$, $\beta_a = 1/3$, $f_M=0.05$, and $\beta_M = 1$, such that the disc mass grows linearly with $M_\mathrm{vir}(z)$ and the scale length grows linearly with $r_\mathrm{vir}(z)$, in good agreement with both empirical constraints \citep{Kravtsov2013} and simulation results \citep{Jiang2019}. The disc properties at $z=0$ are $a_\rmd \approx 3.6$ kpc, $b_\rmd \approx 0.3$ kpc (i.e., a relatively thin disc), and $M_\rmd = 5 \times 10^{10}\, h^{-1} M_\odot$, reminiscent of the Milky Way. The parametrization chosen is sufficiently flexible to enable us to study the impact of the disc growth history and structure on the subhalo population by simply varying $f_a$, $\beta_a$, $f_M$, $\beta_M$, and $b_\rmd/a_\rmd$. As we show in Section~\ref{ssec:params}, though, the results are relatively insensitive to all but the final disc mass ($f_M$). We emphasize that the total mass of the halo--disc system enclosed within $r_\mathrm{vir}(z)$ sums to $M_\mathrm{vir}(z)$. In order to achieve this, we simply multiply the host \citetalias{Navarro.etal.97} density profile by $[M_\mathrm{vir}(z) - M_\rmd(z)]/ M_\mathrm{vir}(z)$.

The merger tree resolution, which sets the lower limit on the subhalo accretion mass, corresponds to $m_\mathrm{res} \geq 10^{8}\, h^{-1} M_\odot$. All subhaloes are evolved until their mass falls below $m = 10^{-5} m_\mathrm{acc}$. All results are averaged over 10,000 host halo merger tree realizations. Specifically, we use the same set of merger trees throughout in order to isolate the differential impact of the disc from the effect of assembly history variation.

In order to strengthen the performance of our mass-loss model with composite halo--disc hosts, we make two changes to \texttt{SatGen} relative to its specification in \citet{Jiang2021} and \citet{Green2021}. Since we are working with an axisymmetric potential, we make the following substitution for the tidal radius definition:
\begin{equation}
    l_\rmt = \Bigg[\frac{m(<l_\rmt)/M(<r)}{2 + \frac{\Omega^2 r^3}{GM(<r)} - \frac{\rmd \ln M}{\rmd \ln r}|_r}\Bigg]^{1/3} \Rightarrow l_\rmt = \Bigg[\frac{Gm(<l_\rmt)}{\Omega^2 - \frac{\rmd^2 \Phi}{\rmd\, r^2}|_{\boldsymbol{r}}}\Bigg]^{1/3} .
\end{equation}
Here, $\Omega$ is the instantaneous angular velocity of the subhalo and $\Phi$ is the gravitational potential of the host halo--disc system. Note that these two definitions of $l_\rmt$, introduced by \citet{King.62}, are identical when the host potential is spherically symmetric. However, the definition on the right is more general, as it does not depend on spherical averages of the host properties. To wit, the radial derivative of an axisymmetric potential is $\frac{\rmd \Phi}{\rmd r} = \frac{\partial \Phi}{\partial R}\frac{\rmd R}{\rmd r} + \frac{\partial \Phi}{\partial z}\frac{\rmd z}{\rmd r}$. Hence, for spherical hosts, the mass-loss model of equation~\eqref{eqn:massloss}, which was calibrated using the definition on the left, remains unchanged after this substitution. The second change is with regards to the mass-loss coefficient, $\alpha$, which is a function of the ratio between the host and subhalo concentrations \citep[see][]{Green2021}. In order to account for the modified mass distribution in the presence of a disc, we modify the host concentration used to compute $\alpha$ according to the following procedure. We define a new scale radius, $r_\mathrm{s,d}$, as the radius within which the enclosed mass of the combined halo--disc system is the same as that which is enclosed within the scale radius of the \citetalias{Navarro.etal.97} halo-only host. The modified host concentration is simply $r_\mathrm{vir}/r_\mathrm{s,d}$, which is somewhat larger than the halo-only concentration due to the compactness of the disc. Note that, unlike \citet{Jiang2021}, we omit the effect of adiabatic contraction on the host concentration since we are interested in studying \textit{relative} disc-driven subhalo depletion. However, we emphasize that a baryon-driven increase in the DM concentration is degenerate with an increased disc mass, which we demonstrate in Section~\ref{ssec:ac}.

Since we calibrated our mass-loss model using the \textit{DASH} simulations, which only include spherical \citetalias{Navarro.etal.97} host haloes, we must verify that the prescription remains valid for subhaloes that evolve in the combined presence of a \citetalias{Navarro.etal.97} halo and a \citetalias{MN1975} disc. To this end, we run an additional set of idealized simulations and compare the predictions of our mass-loss model (with the substituted $l_\rmt$ definition and modified host concentration definition) to the mass trajectories of the simulated subhaloes. The simulation methods and model comparison results are described in Appendix~\ref{app:sims}. In summary, we find that, after making the two aforementioned modifications, the model generalizes well to combined halo--disc potentials and, thus, we proceed to use this modified version of \texttt{SatGen} in this work.

\section{Results}\label{sec:results}

\subsection{Subhalo mass functions}\label{ssec:shmf}

\begin{figure*}
    \centering
    \includegraphics[width=\textwidth]{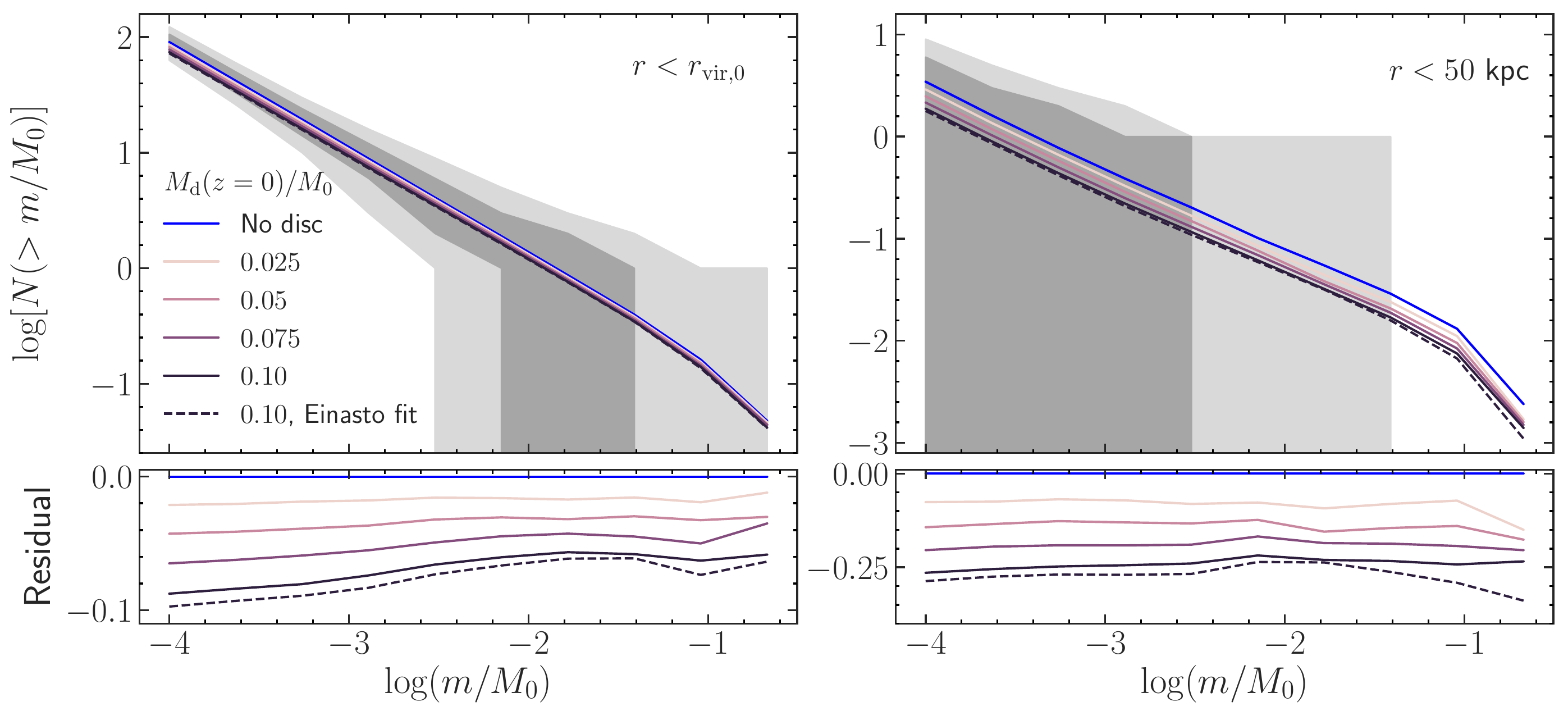}
    \caption{The subhalo mass function, which includes subhaloes of all orders with a $z = 0$ host-centric radius of ({\it left}) $r < r_\mathrm{vir,0}$ or ({\it right}) $r < 50$ kpc. The curves denote the mean SHMF taken over 10,000 trees whereas the dark and light shaded regions correspond to the ${16-84}$ and ${2.5-97.5}$ percentiles, respectively, of the individual (halo-only) trees. The different curves illustrate the dependence of the final disc mass on the subhalo population. The suppression of the subhalo abundance is proportional to the disc mass and is much larger in the halo centre. The dashed lines correspond to the replacement of the $f_M = 0.1$ disc with an \citet{Einasto1965} sphere that has a nearly equivalent enclosed mass profile. The agreement between these dashed lines and the corresponding $f_M = 0.1$ disc curves demonstrates the insignificance of ``disc shocking''.}
    \label{fig:shmfs}
\end{figure*}

We begin by assessing the impact of the disc on the abundance of subhaloes as a function of their mass. Fig.~\ref{fig:shmfs} plots the cumulative subhalo mass functions (SHMF), $N({> m/M_0})$, for the disc-less case (blue lines) as well as for several disc configurations, each with our fiducial $f_a$, $\beta_a$, and $\beta_M$, but with different values of the disc mass fraction, $f_M$, as indicated. The range of disc mass fractions covered ($f_M \in [0.025, 0.05, 0.075, 0.1]$) is motivated by estimates of the stellar mass--halo mass relation \citep[see e.g.,][]{Moster.etal.10, More.etal.11, Behroozi2019}. The dark and light shaded regions denote the ${16-84}$ and ${2.5-97.5}$ percentile intervals, respectively, of the individual SHMFs in the halo-only (i.e., no-disc) case, highlighting the typical halo-to-halo variance. The SHMFs in the left panel include subhaloes of all orders\footnote{A subhalo of order $n$ is hosted by a (sub)halo of order $n-1$, with host haloes corresponding to order $0$.} that have an instantaneous host-centric radius at $z=0$ of $r < r_\mathrm{vir,0}$. In order to emphasize the pronounced effect of the disc in the halo centre, the SHMFs in the right panel are restricted to subhaloes with $r < 50$ kpc. Clearly, the disc results in a suppression of the subhalo abundance that is proportional to the disc mass fraction, $f_M$. However, the halo-to-halo variance in the SHMF is dramatically larger than the difference between the mean SHMFs from the various disc configurations \citep[this finding is also present throughout the results of][]{Jiang2021}. The most massive disc, with $f_M = 0.1$, results in a ${\sim}0.09$ dex (18\%) suppression in the $r < r_\mathrm{vir,0}$ SHMF at the low-mass end, which decreases slightly with increasing $m/M_0$. Our results are in excellent agreement with the idealized simulations of \citet{Errani2017}, who report a 20\% suppression in the SHMF of a cuspy Milky Way-mass halo due to the presence of a disc with $f_M = 0.1$ (note that they also find the effect to be reduced at the high-mass end of the SHMF). The factor of suppression due to the disc is greatly increased when we restrict the SHMF to subhaloes with $r < 50$ kpc. For example, the $f_M = 0.1$ disc drives a ${\sim}0.25$ dex (${\sim} 44\%$) decrease in subhaloes within 50 kpc, which is consistent with \citet{DOnghia2010}. Our mean results are also in excellent agreement with \citet{Jiang2021}, indicating an overall insensitivity to our differing subhalo tidal evolution models.\footnote{Note that \citet{Jiang2021} also account for the fact that the subhalo density profile may be affected by baryons prior to infall. However, this only affects the small fraction of subhaloes that host bright satellites, the analysis of which is beyond the scope of this study.}

We find that the slope of the ${r < r_\mathrm{vir,0}}$ SHMF changes very little with $f_M$, ranging from $-0.91$ in the disc-less case to $-0.89$ in the $f_M = 0.1$ configuration. The same is true of the $r < 50$ kpc SHMF, which has a slope of $-0.81$ in the disc-less case and $-0.79$ when $f_M = 0.1$. Taken together, these results imply that mass segregation \citep[or the lack thereof, cf.][]{vdBosch.etal.16} is not greatly impacted by the disc. However, because of the slight $f_M$-dependence of the slope, the ${r < r_\mathrm{vir,0}}$ SHMF residuals exhibit a small amount of mass-dependence, especially for the larger $f_M$. In order to gauge how this plays out at the low mass end (i.e., $\log(m/M_0) < -4$), we generated 2,000 merger trees with enhanced resolution ($m_\mathrm{acc} \geq 10^7 \, h^{-1}M_\odot$) and evolved the subhaloes using each of the five configurations introduced in Fig.~\ref{fig:shmfs}. We report that the residuals for $10^{-5} \leq m/M_0 \leq 10^{-4}$ flatten off and are consistent with their values at $m/M_0 = 10^{-4}$. This convergence of the slopes in the low-$m/M_0$ limit indicates that the additional dynamical friction due to the disc, which only impacts more massive subhaloes, is the most likely cause of the minor $f_M$-dependence of the SHMF slope.

In order to aid our evaluation of the significance of ``disc shocking'', we consider a case where the \citetalias{MN1975} disc is replaced by a spherical component with a nearly equivalent spherically enclosed mass profile. Specifically, we fit the $M(<r)$ of an \citet{Einasto1965} profile to the $M(<r)$ of a \citetalias{MN1975} disc with $b_\rmd / a_\rmd = 0.08$. Using the notation of \citet[][Appendix A3]{Jiang2021}, the parameters of the resultant \citet{Einasto1965} halo are $M_\mathrm{tot} = M_\rmd$ (i.e., the total mass of the system is unchanged), $n = 2.13$, and $c_2 = 43.7$. The concentration is defined with respect to the $z = 0$ virial mass definition. Using this convention, the mass and size of the spherical substitute grow identically to that of the disc by simply holding $M_{\rm tot} = M_\rmd$. The \citet{Einasto1965} sphere is slightly more centrally concentrated --- its enclosed mass is ${\sim}12\%$ larger than that of the disc at $r \approx 6a_\rmd$, with the $M(<r)$ of the two systems converging as $r$ increases. We evolve the subhaloes in this composite halo--sphere host with $f_M = 0.1$, presenting the resulting SHMF as dashed lines in Fig.~\ref{fig:shmfs}. The results agree quite well with those of the $f_M = 0.1$ disc. The slightly increased central concentration of the spherical replacement appears to drive a minor increase in the overall mass loss relative to the disc. This agreement between the impact of a disc and a spherical replacement is also reported by \citet{Garrison-Kimmel.etal.17}. We elaborate on the implications of this finding in Section~\ref{sec:discuss}.

\subsection{Radial profiles}\label{ssec:radprof}

\begin{figure*}
    \centering
    \includegraphics[width=\textwidth]{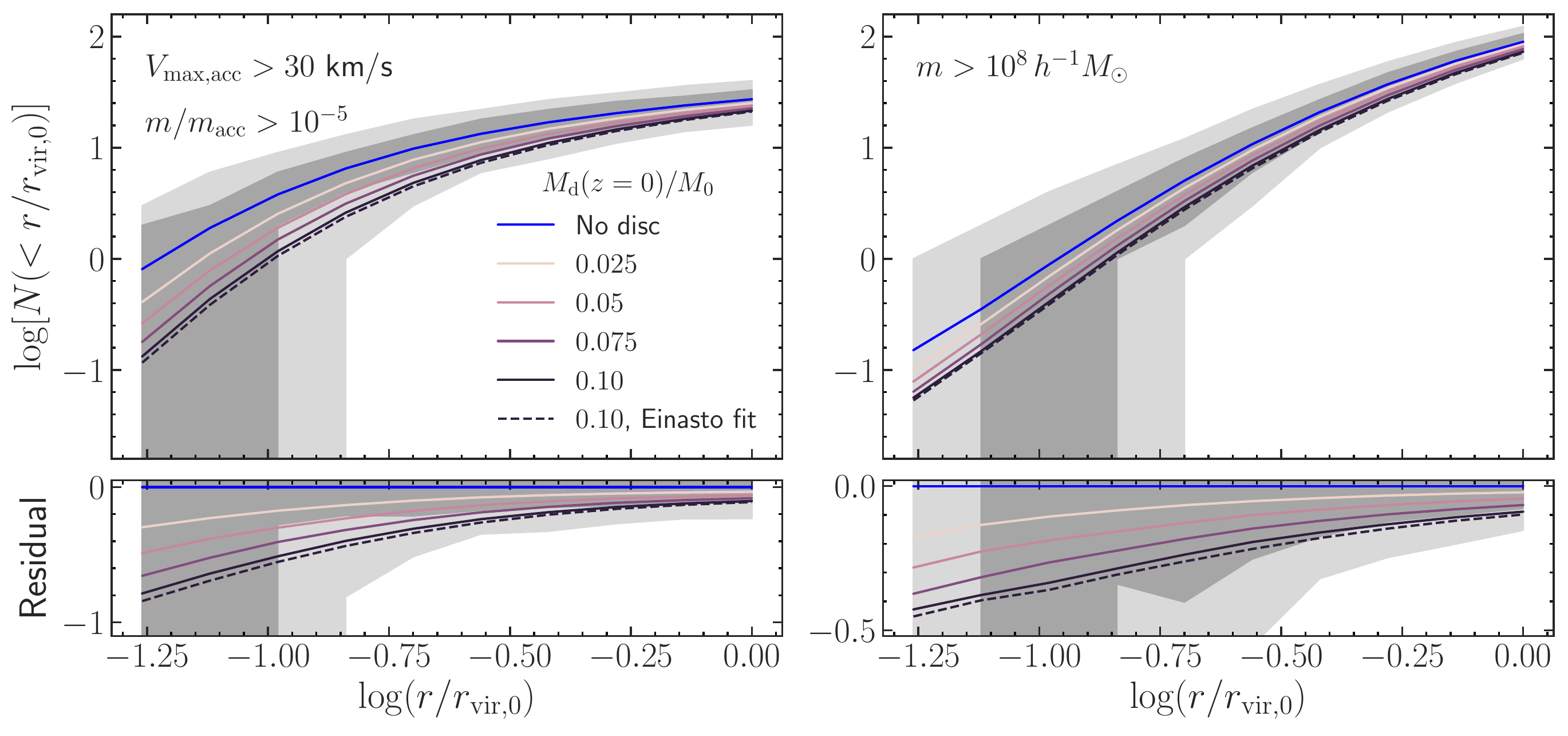}
    \caption{The cumulative radial subhalo abundance profile, which includes subhaloes of all orders with ({\it left}) $V_\mathrm{max,acc} > 30$ km/s and $m/m_\mathrm{acc} > 10^{-5}$ or ({\it right}) $m > 10^{8} \, h^{-1} M_\odot$ at $z = 0$. The meanings of the curves and shaded regions are consistent with those of Fig.~\ref{fig:shmfs}. The disc dramatically reduces the mean abundance of galaxy-hosting subhalo candidates in the halo interior. Nonetheless, the mean curves of all disc configurations lie within the halo-to-halo variance of the disc-less results.}
    \label{fig:radial_abundance}
\end{figure*}

In Fig.~\ref{fig:radial_abundance}, we shift our attention to the cumulative radial subhalo abundance profile, $N(<r/r_\mathrm{vir,0})$. Here, we restrict ourselves to two different subsets of the subhalo population. In the left panel, we only count subhaloes that have a maximum circular velocity at accretion, $V_\mathrm{max,acc}$, that is greater than 30 km/s, which roughly captures the population of subhaloes that could themselves host galaxies. We note that all such subhaloes with $m/m_\mathrm{acc} > 10^{-5}$ are included; however, we find that the results presented throughout this work are qualitatively insensitive to the choice of $m/m_\mathrm{acc}$ used as the cut-off for inclusion. In the right panel, we instead include all subhaloes with instantaneous mass at $z = 0$ of $m > 10^8 \, h^{-1} M_\odot$. Once again, the dark and light shaded regions indicate the 68\% and 95\% halo-to-halo variance intervals in the no-disc case. The mean profile of each disc configuration is again enclosed within the halo-to-halo variance of the disc-less profile. The radial profiles further illustrate the enhanced impact of the disc on subhalo statistics towards the halo centre, as the mean profiles are increasingly suppressed with decreasing $r/r_\mathrm{vir,0}$. This effect is strongest on the population with $V_\mathrm{max,acc} > 30$ km/s, which is reduced by roughly 0.7 dex ($\approx 80\%$) within 20 kpc of the halo centre when $f_M = 0.1$. In comparison to all subhaloes with $m > 10^{8} \, h^{-1} M_\odot$, the population with $V_\mathrm{max,acc} > 30$ km/s is composed of a larger fraction of initially massive subhaloes, which experience stronger dynamical friction and, thus, enhanced tidal stripping. We find that the fractional impact of the disc on the radial profiles of the two simulated systems studied by \citet{Garrison-Kimmel.etal.17} is within the halo-to-halo variance of our log-residuals \citep[see also][]{Jiang2021}. Once again, the replacement of the disc with a nearly equivalent \citet{Einasto1965} sphere results in a radial profile that agrees exquisitely well with that of the $f_M = 0.1$ disc.

\subsection{Enhanced tidal stripping}\label{ssec:strip}

\begin{figure*}
    \centering
    \includegraphics[width=\textwidth]{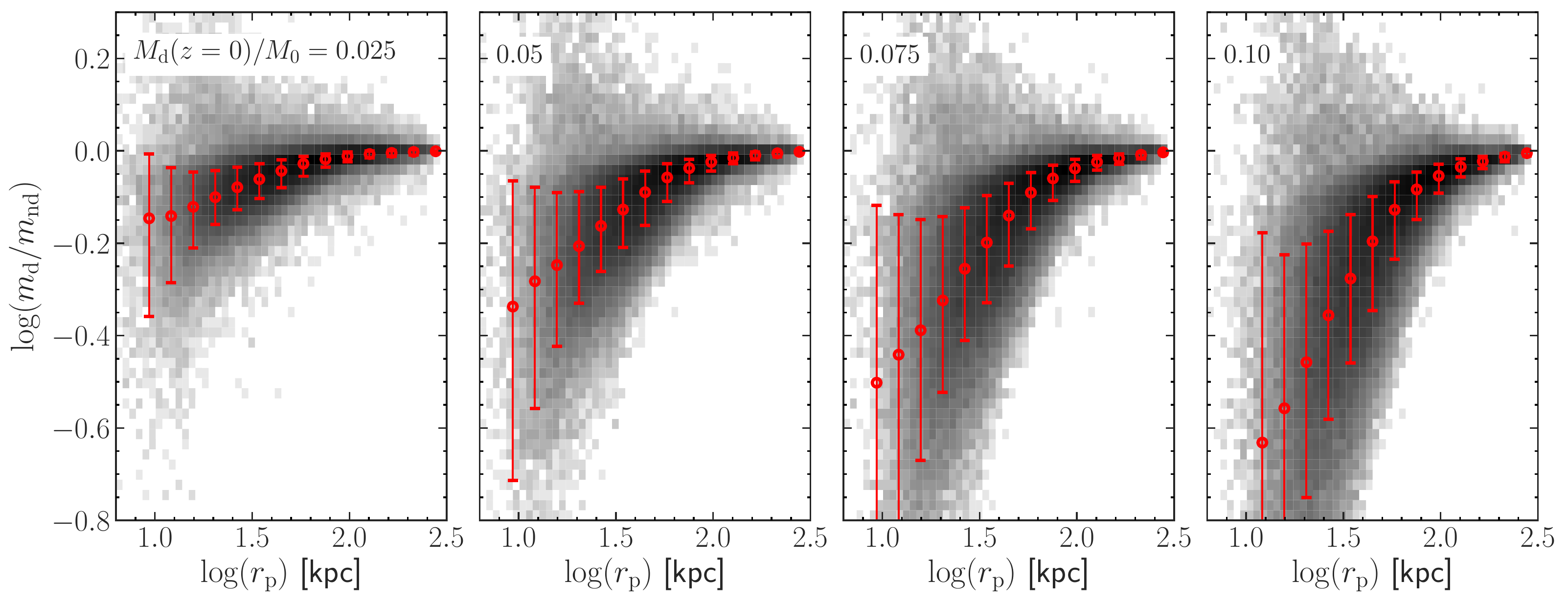}
    \caption{The pericentric radius-dependent impact of a disc on the $z = 0$ mass of individual subhaloes. Each panel denotes a different final disc mass, with the remaining disc parameters held to their fiducial values. The most recent pericentric radius, $r_\rmp$, is measured directly from the \texttt{SatGen} outputs of the halo-only configuration. First-order subhaloes with $V_\mathrm{max, acc} > 30$ km/s, $r < r_\mathrm{vir,0}$, and $m/m_\mathrm{acc} > 10^{-5}$ in both the halo-only and halo--disc configuration that have experienced at least one pericentric passage are included. Here, $m_\rmd / m_\mathrm{nd}$ denotes the ratio between the mass of a particular subhalo in the halo--disc and no-disc configuration. The density heatmap colour is presented on a logarithmic scale. The red circles denote the median $m_\rmd / m_\mathrm{nd}$ in each $r_\rmp$ bin, whereas the error bars correspond to the 68\% interval of halo-to-halo variance. The disc enhances the mass loss of subhaloes that pass near the halo centre while having little impact on those with large-$r_\rmp$ orbits.}
    \label{fig:mass_reduction}
\end{figure*}

By using the same merger tree realizations for all of the halo--disc configurations, \texttt{SatGen} enables us to directly assess the impact of the disc on \textit{individual} subhaloes. In Fig.~\ref{fig:mass_reduction}, we use a log-density heatmap to show the distribution of changes in subhalo mass relative to the no-disc (nd) configuration, expressed via the ratio $m_\rmd / m_\mathrm{nd}$, as a function of the most recent orbital pericentric radius, $r_\rmp$. Each panel corresponds to a different final disc mass, as indicated. We also plot the median $m_\rmd / m_\mathrm{nd}$ in each $r_\rmp$ bin in order to better highlight the trend. The $r_\rmp$ are measured directly from the subhalo position data stored in the \texttt{SatGen} snapshots. For the purpose of this plot, we measure the $r_\rmp$ of each subhalo from the disc-less configuration. However, we acknowledge that the $r_\rmp$ of each subhalo is slightly reduced in the presence of a disc,\footnote{For example, the median (top 1\%) reduction in $r_\rmp$ due to the $f_M=0.1$ disc is 0.004 dex (0.045 dex).} an effect which itself drives a minor enhancement in mass loss. The population of subhaloes included in this analysis have $V_\mathrm{max,acc} > 30$ km/s, are first-order at $z=0$, have $r < r_\mathrm{vir,0}$ and $m/m_\mathrm{acc} > 10^{-5}$ at $z=0$ in both the halo-only and halo--disc configuration, and must have experienced at least one pericentric passage. The figure clearly demonstrates that subhaloes that pass closer to the halo centre experience greater mass loss due to the enhanced central density of the halo--disc system than those that are confined to the halo outskirts. While the median $m_\rmd / m_\mathrm{nd}$ begins to deviate from unity for $r_\rmp \lesssim 50-75$ kpc in all cases, there is a minor indication that the influence of the disc extends out to further radii as its mass is increased. At $r_\rmp \approx 15$ kpc, a disc with $f_M = 0.025$ ($0.1$) drives an additional ${\sim}25\%$ ($70\%$) loss of subhalo mass on the median. These results are consistent with a similar analysis by \citet{Jiang2021}, although we emphasize that we have used a rather different subhalo selection function in this work. Note that the introduction of the disc slightly changes the orbital period, and hence the orbital phase at $z = 0$, of all subhaloes --- this effect results in a small fraction of cases where the subhalo mass is actually larger than its counterpart in the disc-less realization at $z = 0$.

\begin{figure*}
    \centering
    \includegraphics[width=\textwidth]{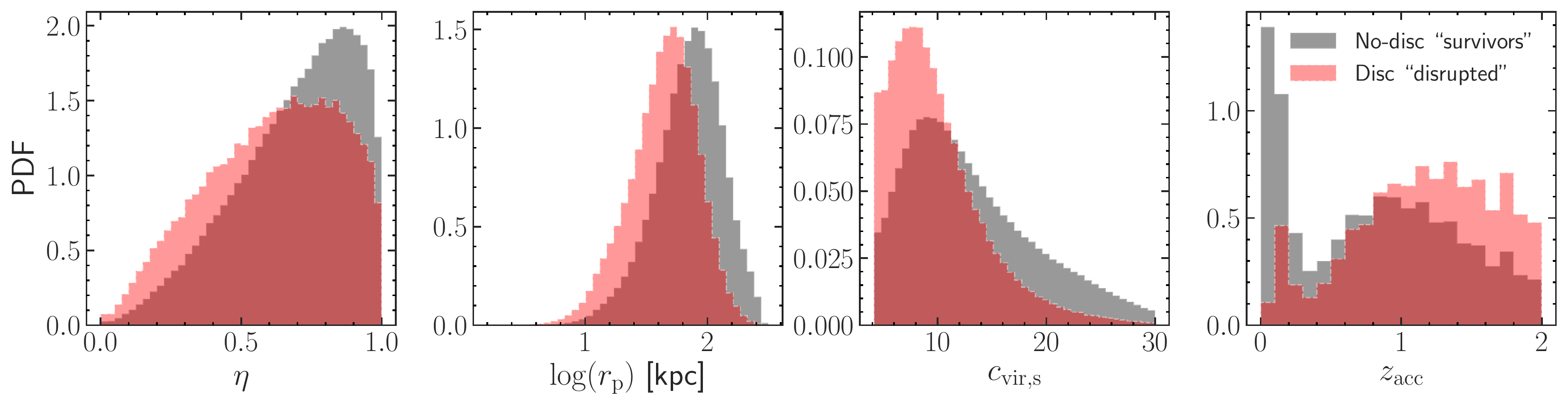}
    \caption{From left to right, the normalized distributions of the instantaneous orbital circularity, $\eta$, radius of first pericentric passage, $r_\rmp$, subhalo concentration at accretion, $c_\mathrm{vir,s}$, and redshift of accretion, $z_\mathrm{acc}$, for two groups of subhaloes in the no-disc configuration. The ``survivors'' group consists of all first-order subhaloes with $m/M_0 > 10^{-4}$ and $r < r_\mathrm{vir,0}$ at $z = 0$ in the no-disc configuration. The ``disrupted'' group is the subset of the no-disc ``survivors'' that have ``disrupted'' (i.e., $m / M_0 < 10^{-4}$ at $z = 0$) in the composite halo--disc configuration with $f_M = 0.1$. The subhaloes that do not survive the additional disc-driven mass loss tend to be on more radial orbits that penetrate more closely into the host centre (smaller mean $\eta$ and $r_\rmp$), are less centrally concentrated, and are typically accreted earlier than the no-disc ``survivors''.}
    \label{fig:disrupted}
\end{figure*}

In simulation-based subhalo studies, the particle resolution imposes a fixed lower mass limit on the subhalo population. An unfortunate limitation of this approach is that a subhalo is typically inferred to have been ``disrupted'' after its mass falls below this limit. We have argued in previous studies \citep[][]{vdBosch.etal.2018b, Green2021} that much of this disruption is not physical and is instead a consequence of the simulation mass limit and artificially enhanced by runaway numerical instabilities. Hence, it is instructive to impose a fixed mass limit on the \texttt{SatGen} results in order to study the properties of subhaloes whose status as ``disrupted'' can be specifically traced to the presence of a disc. Here, we define the set of no-disc ``survivors'' to be all first-order subhaloes with $m/M_0 > 10^{-4}$ and $r < r_\mathrm{vir,0}$ at $z = 0$ in the no-disc realization. The ``disrupted'' group is the subset of the ``survivors'' that instead have $m / M_0 < 10^{-4}$ at $z = 0$ in the halo--disc realization with $f_M = 0.1$ (i.e., the disruption of these subhaloes can be attributed to the presence of the disc). In Fig.~\ref{fig:disrupted}, we present the normalized distributions of several orbital and accretion properties of subhaloes in these two groups. Subhaloes that are most vulnerable to additional disc-driven mass loss, and hence would be preferentially ``disrupted'' by the disc in a simulation, are simply those on more radial orbits \citep[i.e., smaller circularity, $\eta$, as defined in][]{Wetzel.11} that pass closer to the halo centre (smaller $r_\rmp$), are less centrally concentrated at accretion (smaller $c_\mathrm{vir,s}$), and have undergone tidal evolution for a longer period of time (larger $z_\mathrm{acc}$). We expand on the implications of these relatively intuitive findings in the discussion of ``disc shocking'' (Section~\ref{sec:discuss}). Note that a consequence of the preferential ``disruption'' of radially orbiting subhaloes due to the introduction of the disc is a minor change in the subhalo velocity anisotropy towards more circular orbits. The velocity anisotropy profiles predicted by \texttt{SatGen} will be the focus of an upcoming follow-up study.

\subsection{Azimuthal bias of subhaloes}\label{ssec:angles}

It is well known that satellite galaxies are preferentially distributed along the major axis of their central host galaxy \citep[e.g.,][]{Brainerd.05, Yang.etal.06, Azzaro.etal.07}. This ``azimuthal bias'' is typically interpreted as implying that central galaxies are aligned with their host haloes. In particular, numerous studies have pointed out that subhaloes in DM simulations are preferentially distributed along the major axis of their host halo \citep[e.g.,][]{Knebe.etal.04, Libeskind.etal.05, Zentner.etal.05b}. Although a small part of this alignment can be attributed to the preferred direction of subhalo accretion along large-scale filaments \citep[e.g.,][]{Aubert.etal.04, Faltenbacher.etal.08, Morinaga2020}, it is mainly due to the fact that host haloes themselves are not spherical \citep[e.g.,][]{Wang.etal.05, Agustsson.Brainerd.06, Wang.etal.08}. Hence, as long as central galaxies are roughly aligned with their host haloes, this non-spherical distribution of subhaloes naturally explains the azimuthal bias in the observed distribution of satellite galaxies \citep[e.g.,][]{Agustsson.Brainerd.06, Kang.etal.07}.

However, since the central galaxy influences the tidal evolution of the subhaloes, an alternative explanation for the azimuthal bias may be that the central (disc) galaxy preferentially destroys satellite galaxies along more polar-inclined orbits. This would introduce deviations from azimuthal symmetry even when the host halo is spherical and subhaloes are accreted isotropically (as in \texttt{SatGen}).

\begin{figure}
    \centering
    \includegraphics[width=0.47\textwidth]{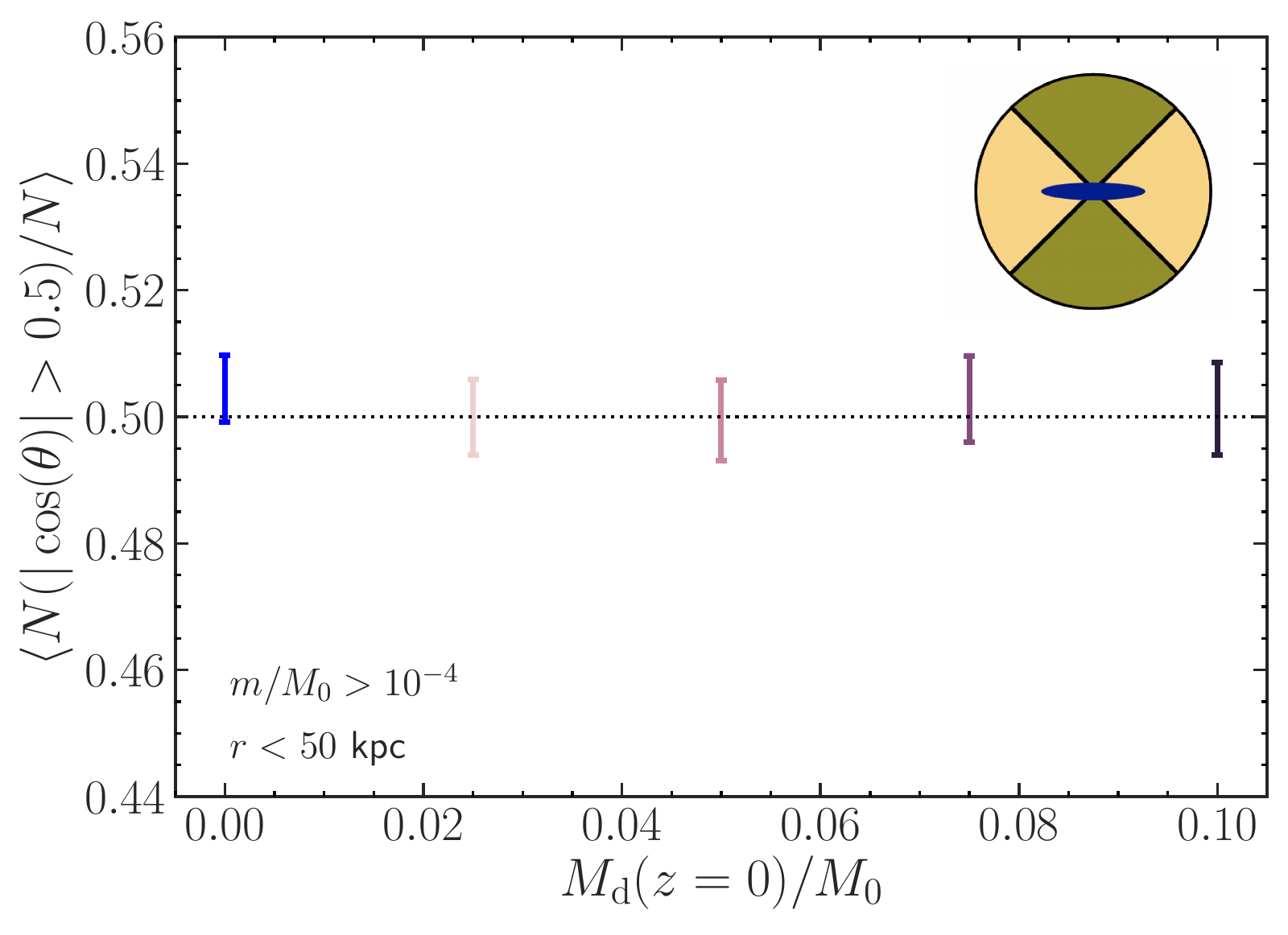}
    \caption{The polar subhalo fraction as a function of the disc mass fraction. The sample includes first-order subhaloes with $m/M_0 > 10^{-4}$ at $z=0$ that lie within $50$ kpc of the host centre and is stacked over the ensemble of hosts. The estimated uncertainty range of the fraction of polar subhaloes, which have host-centric polar angles that satisfy $\abs{\cos(\theta)} > 0.5$ (i.e., those in the green region of the schematic in the upper right), is computed via bootstrap resampling. The error bars denote the 2.5--97.5 percentile intervals of the 2,000 bootstrap estimates of the polar fraction. We find no statistically significant disc-driven azimuthal bias in subhalo positions, regardless of disc mass.}
    \label{fig:angles}
\end{figure}

In order to explore this, we search for an angular bias in our predicted subhalo population as a function of disc mass. In \texttt{SatGen}, the cylindrical $z$-axis is the disc axis of symmetry; hence, a disc-driven bias should manifest itself in the form of an asymmetry in the number of polar subhaloes, with $\abs{\cos(\theta)} > 0.5$, and planar subhaloes, with $\abs{\cos(\theta)} < 0.5$, where $\theta$ is the host-centric spherical polar angle. We restrict our sample to first-order subhaloes with $m/M_0 > 10^{-4}$ at $z=0$ that lie within $50$ kpc of the host centre, but we emphasize that we considered a range of subhalo selection functions and bounding radii and found qualitatively identical results. In order to compute a robust estimate of the mean polar subhalo fraction, $\langle N(\abs{\cos(\theta)} > 0.5) / N \rangle$, and its uncertainty, we stack our subhalo sample over the ensemble of hosts and perform bootstrap resampling. For each disc mass, we generate 2,000 bootstrap estimates of $N(\abs{\cos(\theta)} > 0.5) / N$ and present the 2.5--97.5 percentile intervals in Fig.~\ref{fig:angles}. Note the complete lack of any significant azimuthal bias; on average, there are equal numbers of `polar' and `planar' subhaloes.

\begin{figure}
    \centering
    \includegraphics[width=0.47\textwidth]{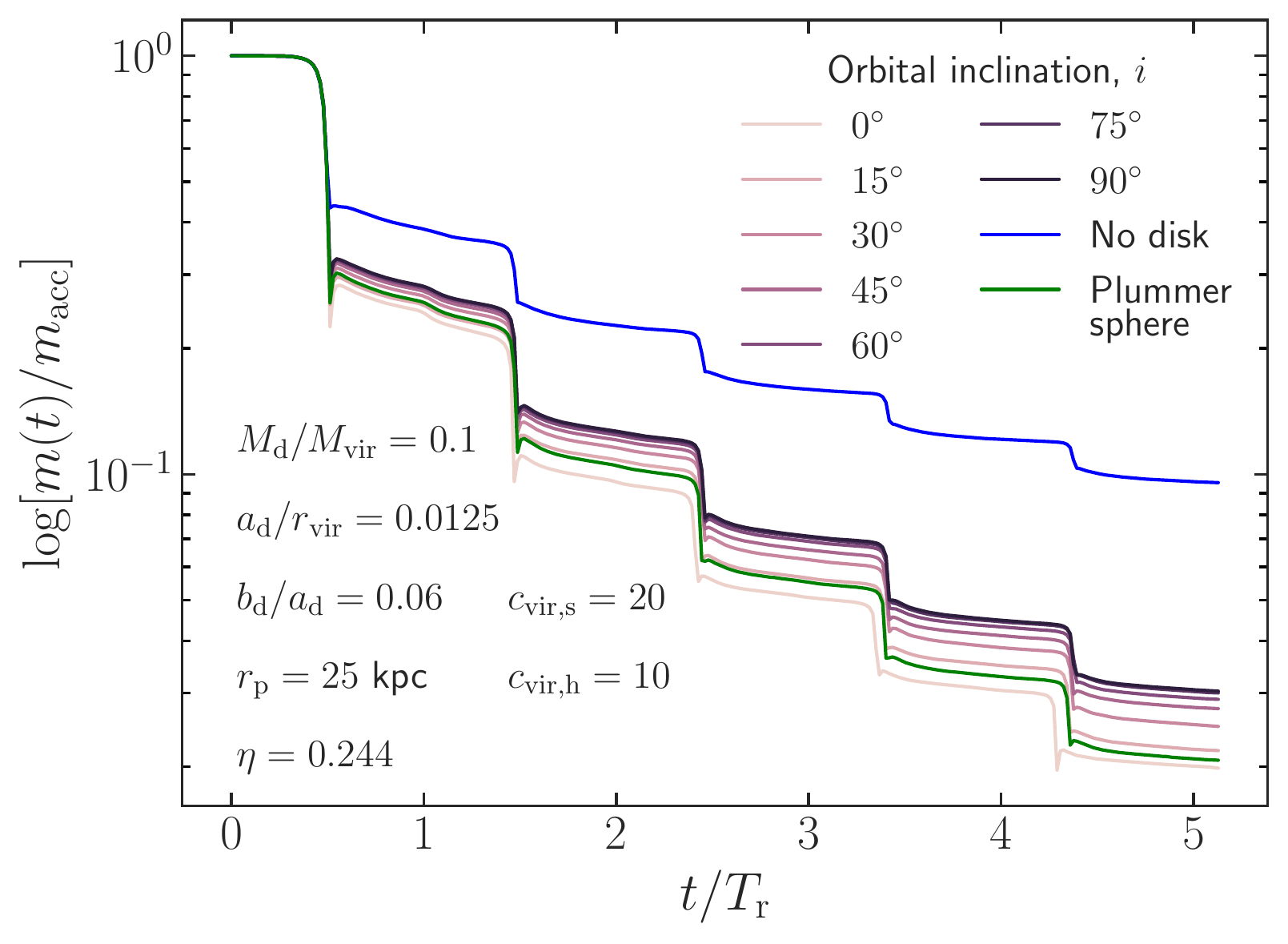}
    \caption{The $m(t)/m_\mathrm{acc}$ trajectories of subhaloes in idealized $N$-body simulations (see Appendix~\ref{app:sims}). The times are normalized by the radial orbital period, $T_r$. The disc mass, disc shape, host and subhalo concentrations, orbital energy, and orbital angular momentum are all held fixed as we vary the orbital inclination angle from $0^{\circ}$ (the orbit is in the plane of the disc) to $90^{\circ}$ (perpendicular to the plane of the disc). As the orbit becomes less inclined, the cumulative mass loss increases, but the overall inclination dependence is weak. Replacing the disc potential with a \citet{Plummer1911} sphere (with the same total mass and a nearly equivalent spherically enclosed mass profile) yields greater mass loss than those of the inclined orbits in the presence of a disc, demonstrating the insignificance of ``disc shocking''.}
    \label{fig:inc}
\end{figure}

We ascribe this lack of azimuthal bias to two effects. First of all, the mass loss of subhaloes depends only weakly on the (polar) angle of incidence between the orientation of the disc and the subhalo orbit. This is demonstrated explicitly in Fig.~\ref{fig:inc}, which shows the $m(t)/m_\mathrm{acc}$ trajectories of subhaloes in idealized $N$-body simulations (see Appendix~\ref{app:sims} for details). Different curves correspond to different orbital inclinations, $i = 90^{\circ} - \theta$, of the initial orbital plane, as indicated, with all other parameters kept fixed. Note that the chosen orbit is highly eccentric, with a small pericentric radius of $r_\rmp = 25$ kpc and orbital circularity of $\eta = 0.244$. For comparison, the blue curve shows the corresponding result in the absence of a central disc and the green curve corresponds to the case where the disc has been replaced by a \citet{Plummer1911} sphere with a nearly equivalent spherically enclosed mass profile, which we discuss in Section~\ref{sec:discuss}.\footnote{The Plummer scale length that yields the best match to the spherically enclosed mass profile of the \citetalias{MN1975} disc (when $b_\rmd / a_\rmd = 0.06$) is $\epsilon \approx 0.92 a_\rmd$. The total mass of the two systems is identical but the \citet{Plummer1911} sphere is slightly more centrally concentrated --- its enclosed mass is ${\sim}10\%$ larger than that of the disc at $r\approx 10a_\rmd$.} Note that more planar orbits (i.e., those with smaller $i$) result in slightly {\it more} mass loss. Hence, if anything, disc-driven disruption should result in a deficit of planar satellites relative to polar satellites, opposite to the trend seen in observational data. The fact that no azimuthal bias emerges is owed to the fact that in an axisymmetric potential the subhalo is not confined to an orbital plane; unlike an orbit in a spherical potential, its polar angle evolves with time. This washes out the weak dependence on the (initial) inclination seen in Fig.~\ref{fig:inc}.

To summarize, we conclude that the angular bias observed in the azimuthal distribution of satellite galaxies does not have its origin in a disc-driven preferential disruption of subhaloes along more polar-inclined orbits. Rather, it is simply due to the central galaxy being aligned with the moment of inertia of the non-spherical host halo combined with the existence of a preferred direction of subhalo accretion due to large-scale filaments.

\subsection{Dependence on disc parameters}\label{ssec:params}

\begin{figure*}
    \centering
    \includegraphics[width=\textwidth]{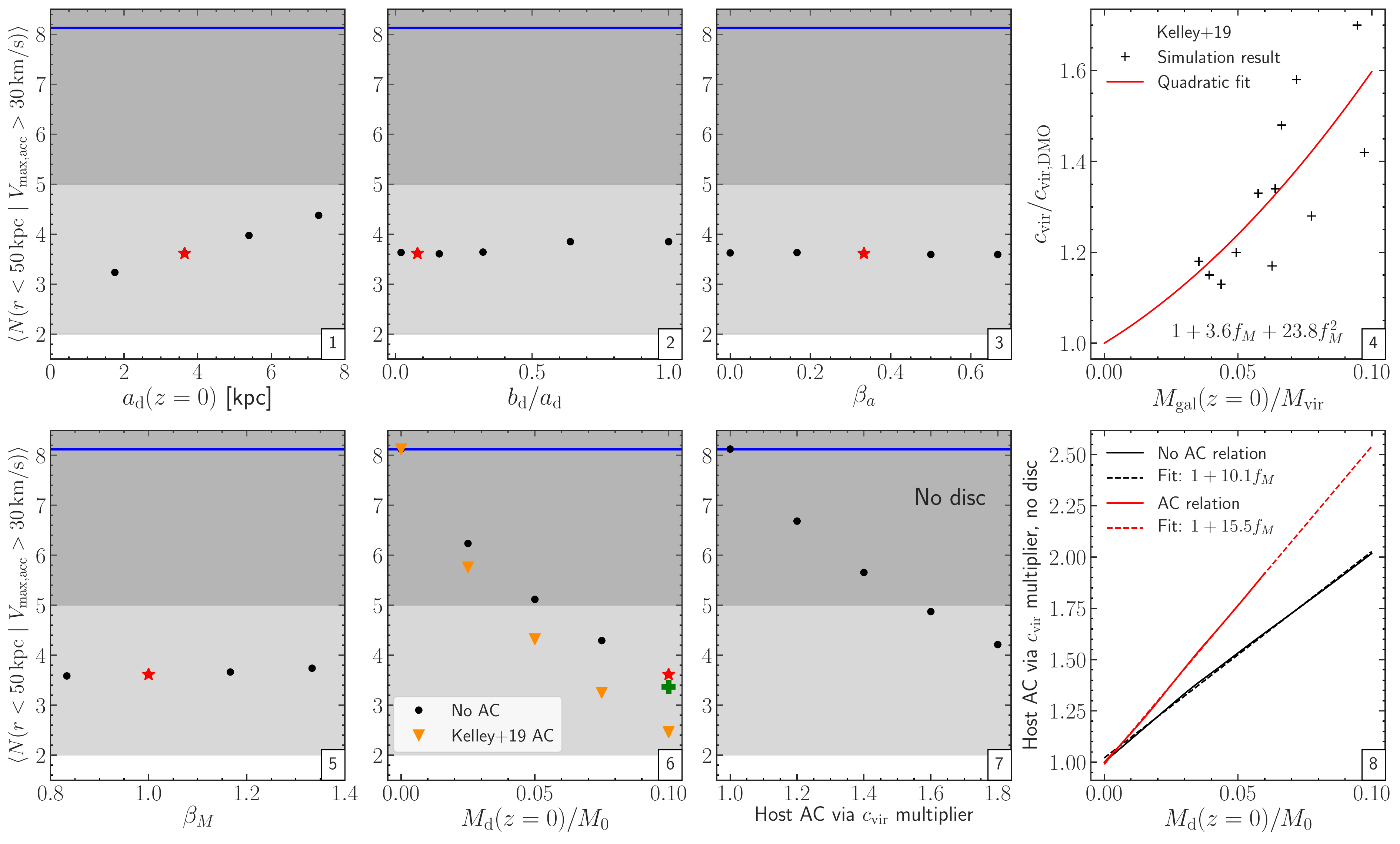}
    \caption{The mean abundance of subhaloes with $V_\mathrm{max,acc} > 30$ km/s and $m/m_\mathrm{acc} > 10^{-5}$ enclosed within 50 kpc of the halo centre at $z = 0$, averaged over all 10,000 merger trees. In panels 1, 2, 3, 5, and 6, we vary one disc parameter, which includes (i) the disc scale length, $a_\rmd (z = 0)$, which is set by $f_a$, (ii) the disc scale height, which is expressed as a fraction of the scale length, $b_\rmd / a_\rmd$, (iii) the power law slope of $a_\rmd (z)$, $\beta_a$, (iv) the power law slope of $M_\rmd (z)$, $\beta_M$, and (v) the disc mass fraction, $M_\rmd(z = 0) / M_0$, which is set by $f_M$. The red star corresponds to the baseline disc configuration ($f_a = 0.0125$, $\beta_a = 1/3$, $f_M=0.10$, $\beta_M = 1$, and $b_\rmd / a_\rmd = 0.08$) around which the parameters are varied. The green cross indicates the spherical \citet{Einasto1965} substitute for the baseline disc. The blue line denotes the $\langle N({r< 50} \, \mathrm{kpc}) \rangle$ of the halo-only configuration, which is surrounded by its 68\% and 95\% halo-to-halo variance intervals (dark and light shaded regions, respectively). Consistent with \citet{Garrison-Kimmel.etal.17}, only the disc mass has a strong impact on the subhalo statistics. Panel 4 displays the relationship (and corresponding quadratic fit) between the mass fraction of a central galactic potential and the corresponding boost in the \citetalias{Navarro.etal.97} concentration of the host due to adiabatic contraction seen in the cosmological simulations of \citet{Kelley2019}. This host contraction due to the disc is accounted for in the orange triangles in panel 6. The impact of an increased host concentration in the \textit{absence} of a disc is shown in panel 7. The degeneracy between host contraction (without a disc) and a disc potential (both with and without adiabatic contraction) is demonstrated in panel 8 --- increasing the disc mass fraction by 0.01 and ignoring (accounting for) adiabatic contraction due to the disc has the same effect on $\langle N({r< 50} \, \mathrm{kpc}) \rangle$ as that of a 10\% (15.5\%) increase in the host concentration.}
    \label{fig:disc_params}
\end{figure*}

In all previous results, we have studied disc configurations with various final masses (controlled by $f_M$) but with only the fiducial $f_a$, $\beta_a$, $\beta_M$, and $b_\rmd / a_\rmd$. In Fig.~\ref{fig:disc_params}, we use a summary statistic to demonstrate that our results are insensitive to these other parameters, which control the disc growth and size. For each disc configuration, we compute the mean number of subhaloes (with $V_\mathrm{max,acc} > 30$ km/s and $m/m_\mathrm{acc} > 10^{-5}$) enclosed within 50 kpc of the halo centre at $z=0$ (averaged over all 10,000 trees), which we denote $\langle N({r< 50} \, \mathrm{kpc}) \rangle$. In panels 1, 2, 3, 5, and 6, we vary one of the disc parameters, fixing the other four to a set of baseline values ($f_a = 0.0125$, $\beta_a = 1/3$, $f_M=0.1$, $\beta_M = 1$, and $b_\rmd/a_\rmd = 0.08$). Note that we use the large $f_M = 0.1$ for our baseline in order to enhance the sensitivity of our results to the other disc parameters. We explore the impact of adiabatic contraction of the host halo in the remaining three panels, which we discuss in Section~\ref{ssec:ac}. For comparison, the horizontal lines indicate the $\langle N({r< 50} \, \mathrm{kpc}) \rangle$ of the no-disc configuration, while the gray shaded regions mark the corresponding 68\% and 95\% halo-to-halo variance intervals. Once again, the mean result lies within the disc-less halo-to-halo variance for every disc configuration studied. Clearly, the mean subhalo abundance within 50 kpc is minimally impacted by the disc scale height and the rate at which the disc grows (both in physical size and mass) relative to the host halo. It is only slightly sensitive to the disc scale length; a more compact disc suppresses more subhaloes. However, the disc mass is the only parameter that has a strong effect on $\langle N({r< 50} \, \mathrm{kpc}) \rangle$ --- the mean abundance drops by 23\% (56\%) relative to the disc-less case when $f_M = 0.025$ ($0.1$). For comparison, the replacement of the $f_M = 0.1$ disc with an \citet{Einasto1965} sphere drives a 59\% suppression in $\langle N({r< 50} \, \mathrm{kpc}) \rangle$ relative to the disc-less case. This finding is in excellent agreement with the cosmological simulation study of \citet{Garrison-Kimmel.etal.17}, who embed a variety of different disc potentials into their host haloes, finding that only the total mass of the disc has a significant impact on the resulting subhalo statistics.

\subsection{Adiabatic contraction of the host}\label{ssec:ac}

Thus far, we have neglected to consider the adiabatic contraction of the host halo due to the formation of the galactic disc. This simplification has enabled us to assess the relative impact of a disc potential on the subhalo population while keeping all other properties of the host consistent with its disc-less counterpart. In addition, proper modeling of adiabatic contraction due to the growth of an axisymmetric potential itself remains an open problem. For example, the `standard' adiabatic invariant, $r M(r)$, originally suggested by \citet{Barnes.White.84} and \citet{Blumenthal.etal.86} and often used in modeling disc galaxies and their rotation curves \citep[e.g.,][]{Mo.etal.98, vdBosch.Swaters.01}, is only valid for unrealistic, completely spherical systems in which all particles move on circular orbits. Furthermore, the fact that the scatter in the Tully--Fisher relation is independent of size \citep[e.g.,][]{Courteau.Rix.99, Courteau.etal.07} has been used to argue that disc formation cannot be associated with significant halo contraction \citep[e.g.,][]{Dutton.etal.07}, which might have its origin in non-adiabatic processes operating during disc formation \citep[see e.g.,][]{ElZant.etal.01, Tonini.etal.06}.

Despite these issues, we now proceed to investigate how adding (adiabatic) contraction of the host halo due to the assembly of the disc impacts the subhalo population. Rather than assuming a particular adiabatic invariant, we consider a simplified model of the halo contraction based on the simulation results of \citet{Kelley2019}. These authors run a suite of twelve dark matter-only cosmological zoom-in simulations, each of which is centred on a different Milky Way-like halo. They then re-run each of the simulations with an embedded galactic potential, which grows over time, placed at the centre of each halo. They fit a \citetalias{Navarro.etal.97} profile to each host halo at $z = 0$ in both the halo-only and halo--disc configurations and report the concentrations ($c_\mathrm{vir,DMO}$ and $c_\mathrm{vir}$, respectively). In panel 4 of Fig.~\ref{fig:disc_params}, we plot the ratio, $c_\mathrm{vir}/c_\mathrm{vir,DMO}$, as a function of the fraction of mass in the embedded potential, $M_\mathrm{gal}(z = 0)/M_\mathrm{vir}$. Note that \citet{Kelley2019} model the galaxy as a composite potential that consists of a stellar disc, a gaseous disc, and a stellar bulge; we define $M_\mathrm{gal}$ to be the combined mass of these three systems and compare it directly with the mass of our single-component stellar disc. Clearly, when the central galaxy makes up a larger fraction of the total host mass, the host experiences greater contraction, which corresponds to a larger $c_\mathrm{vir} /c_\mathrm{vir,DMO}$. We fit a quadratic to the relationship, demanding the physical constraint that $c_\mathrm{vir} /c_\mathrm{vir,DMO} = 1$ when $M_\mathrm{gal}(z = 0)/M_\mathrm{vir} = 0$. We use this fit as the basis of our adiabatic contraction model.

We emulate adiabatic contraction using the following approach. Given the $f_M$ of the disc potential of interest, we look up the corresponding $c_\mathrm{vir} /c_\mathrm{vir,DMO}$ using the fit in panel 4 of Fig.~\ref{fig:disc_params}. Our composite host system is exactly the same as before except that the concentration of the \citetalias{Navarro.etal.97} host halo is multiplied by the corresponding value of $c_\mathrm{vir} /c_\mathrm{vir,DMO}$ at all times during the evolution of the subhaloes. For $f_M = 0.05$ ($0.1$), the host concentration is boosted by a factor of 1.24 (1.60). Note that the total mass of the composite system enclosed within $r_\mathrm{vir}$ remains unchanged. In panel 6 of Fig.~\ref{fig:disc_params}, the orange triangles show the mean subhalo abundance within 50 kpc of the host centre, $\langle N({r< 50} \, \mathrm{kpc}) \rangle$ (a good summary of the overall influence of the disc), as a function of $f_M$ when the host concentration is boosted in order to account for adiabatic contraction. These can be directly compared to the case without adiabatic contraction for each $f_M$, denoted by the black circles. As expected, as the disc becomes more massive, the relative impact of the host contraction becomes more significant. Relative to the halo-only host, the $f_M = 0.05$ ($0.1$) disc suppresses $\langle N({r< 50} \, \mathrm{kpc}) \rangle$ by 37\% (56\%) without adiabatic contraction and by 47\% (70\%) when the host concentrations are boosted according to the \citet{Kelley2019} model. Note that the $\langle N({r< 50} \, \mathrm{kpc}) \rangle$ of the most massive disc remains within the halo-to-halo variance of the halo-only configuration even when adiabatic contraction is taken into account, further demonstrating the importance of such variance in subhalo statistics.

In Section~\ref{sec:methods}, we claimed that the impact of adiabatic contraction is degenerate with simply increasing the disc mass. The reason for this is simple: increasing the disc mass increases the central concentration of the host mass, which is the same result as that of adiabatic contraction. In panel 7 of Fig.~\ref{fig:disc_params}, we consider how $\langle N({r< 50} \, \mathrm{kpc}) \rangle$ of the halo-only configuration is suppressed if we boost the host concentration by a constant factor at all times during subhalo evolution. In the absence of a disc, boosting the host concentration by a factor of $1.6$ drives a 40\% reduction in $\langle N({r< 50} \, \mathrm{kpc}) \rangle$. This level of adiabatic contraction is seen in systems with a disc with $f_M = 0.1$, which itself suppresses $\langle N({r< 50} \, \mathrm{kpc}) \rangle$ by 56\% without the concentration boost. Hence, the effect of the disc itself and the adiabatic contraction that it brings about are both of similar importance, although the disc drives slightly more subhalo suppression. Since the host contraction and disc mass are degenerate, we can use both of panels 6 and 7 in Fig.~\ref{fig:disc_params} in order to understand the relationship between disc-driven suppression (both with and without also accounting for adiabatic contraction; in terms of $\langle N({r< 50} \, \mathrm{kpc}) \rangle$) and concentration boost-driven suppression (in the absence of a disc). We interpolate between both the black circles and orange triangles in panel 6 and the black circles in panel 7 and then perform the following: for each value of $M_\rmd(z=0)/M_0$ (separately with and without adiabatic contraction), we find the value of the host $c_\mathrm{vir}$ multiplier (without a disc) that corresponds to the same $\langle N({r< 50} \, \mathrm{kpc}) \rangle$. These two relationships are shown in panel 8 of Fig.~\ref{fig:disc_params} alongside nearly perfect linear fits. In terms of $\langle N({r< 50} \, \mathrm{kpc}) \rangle$, increasing the disc mass fraction by 0.01 and ignoring (accounting for) adiabatic contraction due to the disc has the same effect as a 10\% (15.5\%) increase in the host concentration. Hence, the impact of a disc potential can be roughly emulated by simply making the host halo more concentrated.

\section{Discussion}\label{sec:discuss}

As we have seen, the presence of a central disc galaxy causes a suppression in the abundance of subhaloes. Our \texttt{SatGen}-based results are in good agreement with previous results based on $N$-body simulations \citep[][]{DOnghia2010, Garrison-Kimmel.etal.17, Errani2017, Sawala.etal.17, Kelley2019}, both qualitatively and quantitatively. However, we disagree with \citet{DOnghia2010} and \citet{Garrison-Kimmel.etal.17} with regards to the importance and origin of this disc-induced substructure depletion. In particular, contrary to these previous studies, we argue that the presence of a disc does not cause actual, physical disruption of substructure. Rather, it merely causes enhanced stripping. We emphasize that this is not just a semantic issue; rather, it is the difference between having no substructure within the inner $30$ kpc of the Milky Way and having thousands of low mass subhaloes (a small fraction of which may host a satellite galaxy) with reduced mass compared to a case without a disc.

In particular, \citet{DOnghia2010} argued that subhaloes that pass near a central disc galaxy will be destroyed due to impulsive ``disc shocking''.  Whenever a subhalo passes through the plane of a disc at sufficiently high speed, its internal energy will increase by an amount $\Delta E$ that can be calculated analytically \citep[][]{Ostriker.72, Binney.Tremaine.08}. The crux of the argument made by \citet{DOnghia2010} is that whenever this $\Delta E$ exceeds the binding energy of the subhalo, $\abs{E_\rmb}$, the subhalo is `certain' to be disrupted. They proceed to show that this condition is met by a significant fraction (${\sim}15\%$) of subhaloes in a realistic, MW-like system. 

However, as shown in \citet{vdBosch.etal.2018a}, it is incorrect to assume that $\Delta E > \abs{E_\rmb}$ will result in disruption. What matters is not only the total energy injected but also how that energy is distributed over the constituent particles. Since $\Delta E \propto l^2$, the particles in the outskirts of the subhalo, which need little energy to escape, receive the bulk of the energy injection whereas the particles near the subhalo centre, which need a large amount of energy to escape, receive virtually no energy. As a consequence, subhaloes can actually experience a tidal shock that exceeds many multiples of their binding energy and still survive. In fact, \citet{vdBosch.etal.2018a} show that subhaloes with a \citetalias{Navarro.etal.97} profile that experience a tidal shock with $\Delta E/\abs{E_\rmb} = 1$ ($10$) only lose roughly 20\% (55\%) of their mass.

Another argument against disc shocking is the fact that in most cases it is subdominant to `halo shocking', which is tidal heating due to a high speed pericentric passage with respect to the host halo itself. Indeed, \citet{vdBosch.etal.2018a} showed that the average $\Delta E/\abs{E_\rmb}$ of a subhalo due to its first pericentric passage in a (disc-less) host halo is about ${\sim}1.9$, which is larger than the average $\Delta E/\abs{E_\rmb}$ due to disc shocking. This is consistent with \citet{DOnghia2010}, who showed that halo shocking dominates over disc shocking in a typical MW-like system except for subhaloes with pericentric radii smaller than ${\sim}10$ kpc.

The insignificance of disc shocking is also evident from Fig.~\ref{fig:inc}. If disc shocking were indeed the dominant factor in disc-driven subhalo depletion, then a subhalo on a highly inclined orbit (relative to the disc) should experience more mass loss than a subhalo whose orbit is confined to the plane of the disc. However, Fig.~\ref{fig:inc} demonstrates that the exact opposite trend is seen in our idealized $N$-body simulations (see Appendix~\ref{app:sims}), which is nicely reproduced by our tidal stripping model that is implemented in \texttt{SatGen}. Furthermore, replacing the disc potential with an equivalent \citet{Plummer1911} sphere in an idealized simulation (see Fig.~\ref{fig:inc}) results in greater mass loss than those of the inclined orbits in the presence of a disc. This is consistent with our \texttt{SatGen} results, where Figs.~\ref{fig:shmfs}, \ref{fig:radial_abundance}, and \ref{fig:disc_params} demonstrate that subhalo statistics are generally insensitive to the replacement of the disc with a spherical system that has an equivalent spherically enclosed mass profile.

In fact, our tidal stripping-based mass-loss model successfully reproduces the subhalo mass evolution of the idealized halo-only \textit{DASH} simulations, our new composite halo--disc simulations (Appendix~\ref{app:sims}), and a wide range of results from the \textit{Bolshoi} cosmological simulations \citep{Green2021} \textit{without an explicit prescription for tidal shocking}. The model relies upon the tight empirical relationship between the stripped subhalo density profile and the bound mass fraction, which is independent of the details of how the mass was lost \citep[e.g.,][]{Hayashi.etal.03, Penarrubia2008, Green.vdBosch.19}. The accuracy of our mass-loss model and the strength of the density profile--mass fraction relationship would be substantially reduced if tidal shocking dominated for some orbits (i.e., small $r_\rmp$) and tidal stripping dominated for others (i.e., large $r_\rmp$).

Based on all of these considerations, we conclude that disc shocking plays, at most, a minor role in the substructure suppression caused by a disc potential. Instead, the presence of a disc greatly increases the central mass concentration of the host, which results in an overall increase in tidal stripping that becomes increasingly significant as $r_\rmp$ becomes smaller. This net increase in subhalo mass loss effectively shifts the mean SHMF to the left (see Fig.~\ref{fig:shmfs}). The total subhalo abundance above a particular simulation mass limit is decreased. However, considering the fact that a \citetalias{Navarro.etal.97} subhalo should never \textit{fully} disrupt \citep[see e.g.,][]{vdBosch.etal.2018a, Errani2020, Errani2021}, we emphasize that this reduced abundance is not due to ``disruption'', but is instead a consequence of enhanced mass loss combined with a fixed resolution limit.

\section{Summary}\label{sec:summary}

The demographics of DM substructure depend on both the particle nature of DM and the gravitational interaction between DM and baryons. Hence, in order to understand the dependence on the former, we must be able to properly account for the latter. Much progress has been made towards correctly capturing the manner in which baryons shape the overall DM distribution \citep[e.g.,][]{DOnghia2010, Zolotov.etal.12, Brooks2013, Garrison-Kimmel.etal.17, Sawala.etal.17, Kelley2019}. However, a common limitation of these studies is that they are all based on expensive cosmological simulations, which has limited their ability to consider statistically complete halo samples and properly contextualize results in terms of the corresponding halo-to-halo variance. The primary finding of these works is clear: the presence of a galactic disc suppresses subhalo abundance, an effect that becomes stronger towards the halo centre. Since the suppression also increases with increasing disc mass, properly accounting for this disc-driven subhalo depletion is especially important for Milky Way-mass systems, which sit at the peak of the stellar mass--halo mass relation.

In this paper, we used the \texttt{SatGen} semi-analytical modeling framework to assess the impact of a galactic disc potential on the DM subhalo demographics of MW-like hosts. This method is not impacted by issues related to numerical disruption, which still hamper the results of $N$-body simulations, and allows for the construction of large halo samples, which enables unprecedented statistical power. Using an ensemble of 10,000 merger trees with $M_0 = 10^{12} \, h^{-1} M_\odot$, we generated an equally large sample of evolved subhalo populations using a range of different composite halo--disc potentials. This approach allowed us to isolate the differential influence of the disc by controlling for assembly history variance. Leveraging the computational efficiency of \texttt{SatGen}, we explored a wide range of disc parameter space, spanning the disc mass, size, and formation history. We used the resulting subhalo catalogs to study subhalo mass functions and radial abundance profiles. We also measured the relative impact of the disc on the $z = 0$ mass of individual subhaloes as well as examined whether disc-driven subhalo depletion gives rise to an azimuthal bias in the spatial distribution of the subhalo population. Our most notable findings are summarized as follows:
\begin{itemize}
\item For a disc mass fraction of $f_M = 0.05$, which is a typical value for a Milky Way-size halo, the normalization of the mean SHMF (of subhaloes with $r < r_\mathrm{vir,0}$) is suppressed by $\lesssim 10\%$ relative to the no-disc case. When only considering subhaloes within 50 kpc of the halo centre, the mean SHMF normalization is decreased by ${\sim} 30 \%$. The level of substructure suppression increases with disc mass. However, the mean disc-driven impact on the SHMF is dwarfed by the halo-to-halo variance in all cases.
\item The disc has a considerably larger influence on the subhalo abundance near the halo centre, as evidenced by the mean radial subhalo abundance profiles. For example, the mean abundance of potential galaxy-hosting subhaloes (with $V_\mathrm{max, acc} > 30$ km/s) is suppressed by ${\sim}40\%$ within 50 kpc of the halo centre relative to the no-disc case when $f_M = 0.05$ but is reduced by ${\sim} 60\%$ within 20 kpc. The mean effect of the disc on the radial profile is again eclipsed by the halo-to-halo variance.
\item By tracking individual subhaloes across different host halo--disc configurations, we have shown that the presence of a central disc causes excess subhalo mass loss, the strength of which increases with decreasing pericentric radius. For example, at $r_\rmp \approx 50$ kpc ($20$ kpc), a disc with $f_M = 0.05$ drives an additional ${\sim}15\%$ ($40\%$) loss of subhalo mass on the median.
\item By imposing a fixed mass resolution limit ($m/M_0 > 10^{-4}$), consistent with simulation-based subhalo studies, we analyzed the orbital and accretion properties of subhaloes that survive until $z = 0$ in the absence of a disc but are ``disrupted'' (i.e., their $m$ falls below the mass cut) by $z = 0$ in the composite host halo--disc case. On average, these disc-disrupted subhaloes are found to have smaller $r_\rmp$ and $c_\mathrm{vir,s}$ than the overall sample.
\item The presence of the disc does not cause an azimuthal bias in the spatial distribution of subhaloes for any of the disc masses considered ($f_M \leq 0.1$). Therefore, the observed alignment of satellite galaxies with the orientation of their central host is not driven by the presence of a disc, but is instead an outcome of galaxy-halo alignment in non-spherical haloes.
\item The overall amplitude of disc-driven subhalo depletion is relatively insensitive to the size of the disc and its detailed formation history (both in terms of its size and mass). Rather, it depends almost exclusively on the final mass of the disc. The replacement of the disc with a spherical system, which has a nearly equivalent spherically enclosed mass profile, of the same total mass yields subhalo statistics that are in excellent agreement with the analogous halo--disc configuration.
\item We demonstrated that the impact of a disc potential can be emulated by simply increasing the concentration of the host halo. Increasing the disc mass fraction by 0.01 and ignoring (accounting for) adiabatic contraction due to the disc has the same impact on $\langle N({r< 50} \, \mathrm{kpc}) \rangle$ as boosting the host concentration by 10\% (15.5\%). Adiabatic contraction of the host due to the formation of the galactic disc only has a significant effect on the overall subhalo abundance when the disc mass fraction is large.
\end{itemize}

Overall, our \texttt{SatGen}-based results are in excellent agreement (both qualitatively and quantitatively) with previous results based on $N$-body simulations \citep[][]{DOnghia2010, Garrison-Kimmel.etal.17, Errani2017, Sawala.etal.17, Kelley2019}. However, as discussed in detail in Section~\ref{sec:discuss}, we disagree with the notion promoted by \citet{DOnghia2010} and \citet{Garrison-Kimmel.etal.17} that the disc causes actual disruption of subhaloes via impulsive disc shocking. Rather, the disc simply increases the density in the central region of the halo, which promotes excess mass loss. When this enhanced mass loss results in the subhalo mass dropping below the resolution limit of a numerical simulation, the subhalo appears to have been disrupted; in reality, it would continue to survive with a reduced mass \citep[and should never \textit{fully} disrupt; see e.g.,][]{vdBosch.etal.2018a, Errani2020, Errani2021}. Another new insight that has emerged from this study relates to the overall importance of disc-driven subhalo depletion. By using a large ensemble of merger trees, we were able to demonstrate that the impact of the disc is small compared to the expected halo-to-halo variance, even for the most massive discs considered. Hence, when using the abundance of satellite galaxies or subhaloes in a single system, such as the Milky Way, it is more important to account for halo-to-halo variance than the impact of the central galaxy when making inferences.

\section*{Acknowledgements}

The authors thank Uddipan Banik and Rapha{\" e}l Errani for helpful conversations throughout the development of this work. SBG is supported by the US National Science Foundation Graduate Research Fellowship under Grant No. DGE-1752134. FCvdB is supported by the National Aeronautics and Space Administration through Grant No. 17-ATP17-0028 issued as part of the Astrophysics Theory Program. FJ is supported by the Troesh Fellowship from the California Institute of Technology.

\section*{Data availability}

The updated \texttt{SatGen} library is publicly available on GitHub.\footnote{\href{https://github.com/shergreen/SatGen/}{https://github.com/shergreen/SatGen/}}


\bibliographystyle{mnras}
\bibliography{references_vdb}


\appendix
\section{Idealized simulations}\label{app:sims}

In order to verify that the \citet{Green2021} mass-loss model, which is given by equation~\eqref{eqn:massloss}, can be used to accurately describe subhalo evolution in a combined halo--disc host system, we run a set of idealized simulations that serve as the ground truth for comparison to our model predictions. We use the same procedure as used for the \textit{DASH} simulations \citep[][]{Ogiya2019}. In particular, all simulations are run using the $N$-body code \texttt{OTOO+} \citep{Ogiya2013}, which is a GPU-accelerated tree code. The simulations follow the evolution of a live $N$-body subhalo (initially a \citetalias{Navarro.etal.97} halo composed of $N=10^6$ particles) as it orbits within a static \citetalias{Navarro.etal.97} host halo potential (with initial sub-to-host mass ratio of $m/M=10^{-3}$) for 36 Gyr.

In contrast to \textit{DASH}, the host system in our test suite of simulations is composed of a \citetalias{Navarro.etal.97} halo with an embedded \citetalias{MN1975} disc. The host halo has a concentration of $c_\mathrm{vir,h} = 10$ and the subhalo has an initial concentration of $c_\mathrm{vir,s} = 20$, which is consistent with a typical minor merger with a Milky Way-like host. For simplicity, we only consider orbits with orbital energy equal to that of a circular orbit at the virial radius of the host, which coincides with the peak of the orbital energy distribution of infalling subhaloes in cosmological simulations \citep[e.g.,][]{Li2020}. 

When constructing our simulation suite, we vary several parameters, which control properties of the disc, the inclination of the initial subhalo orbit with respect to the disc, and the radius of orbital pericentre. The fraction of the host mass in the disc is set by $f_M \equiv M_\rmd / M_\mathrm{vir} \in [0.0, 0.02, 0.05, 0.1]$. In order to preserve the total mass enclosed within $r_\mathrm{vir}$, we multiply the host halo density profile by $1-f_M$. The disc scale length is varied over $f_a \equiv a_\rmd / r_\mathrm{vir} \in [0.007, 0.0125, 0.025]$, with $f_a = 0.0125$ corresponding to $a_\rmd \approx 3.5$ kpc for the Milky Way-mass host, while the disc scale height is controlled by varying $b_\rmd / a_\rmd \in [0.02, 0.06, 0.2]$. We consider seven orbital inclinations with $i \in [0, 15, 30, 45, 60, 75, 90]$ degrees, where $i$ is the angle between the orbital plane and the disc (i.e., $i = 0^{\circ}$ results in a subhalo that orbits in the plane of the disc). The final parameter is the orbital angular momentum, which we adjust such that the pericentric radius, $r_\rmp$, of the subhalo orbit \textit{in the no-disc configuration} is equal to $15$, $25$, or $50$ kpc.

Given the orbital energy, angular momentum, and inclination angle, we initialize the subhalo at its apocentre. Note that we use the same initial position and velocity regardless of the disc properties. Hence, the true $r_\rmp$ attained by the subhalo varies slightly with $f_M$, $f_a$, and $b_\rmd/a_\rmd$.

\begin{figure}
    \centering
    \includegraphics[width=0.44\textwidth]{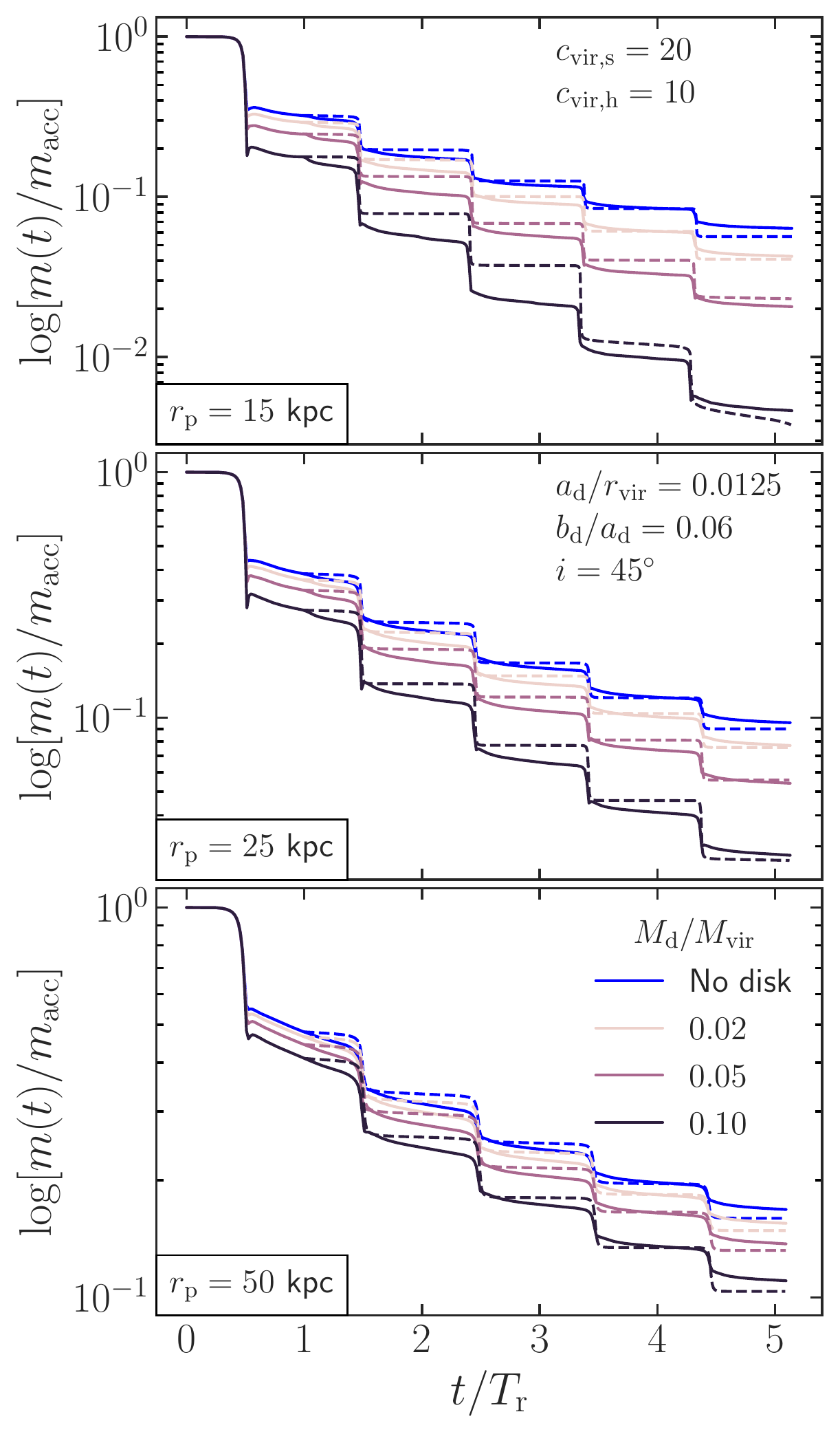}
    \caption{A comparison between the \citet{Green2021} mass-loss model predictions (dashed lines) and the $m(t)/m_\mathrm{acc}$ trajectories of several simulated subhaloes (solid lines). The times are normalized by the radial orbital period, $T_r$. The host and subhalo concentrations, disc shape, orbital inclination, and orbital energy are held fixed. The model performs well over the full range of disc masses and down to small pericentric radii ($r_\rmp$).}
    \label{fig:fbt}
\end{figure}

Thus, our test suite spans $f_M$, $f_a$, $b_\rmd / a_\rmd$, $i$, and $r_\rmp$. We follow the procedure laid out in Section~2.3.2 of \citet{Green2021} to generate mass-loss model predictions and make comparisons to the bound-mass trajectories, $m(t)/m_\mathrm{acc}$, of the simulated subhaloes. The performance of the model is illustrated in Fig.~\ref{fig:fbt}, which compares the simulation and model results for the case of a fiducial MW-like disc shape and an inclined subhalo orbit with $i=45^{\circ}$. Clearly, the model accurately captures the subhalo mass evolution for all $M_\rmd / M_\mathrm{vir}$ and $r_\rmp$ considered. When averaged over the full test suite, we find that our model remains unbiased with relatively low scatter in the log-residuals (less than a factor of two larger than that of the halo-only case) for longer than a Hubble time, indicating that mass evolution error will be subdominant to the halo-to-halo variance. Hence, the mass-loss model successfully captures the additional loss of mass due to the presence of a disc, validating its use in \texttt{SatGen} for the present study.

\bsp
\label{lastpage}
\end{document}